# Recollision induced superradiance of ionized nitrogen molecules

Yi Liu[1], Pengji Ding[1], Guillaume Lambert[1], Aurélien Houard[1],
Vladimir Tikhonchuk[2], and André Mysyrowicz[1], *

1. Laboratoire d'Optique Appliquée, ENSTA ParisTech, CNRS, Ecole polytechnique, Université Paris-Saclay, 828 bd des Maréchaux, 91762 Palaiseau cedex France
2. Centre Lasers Intenses et Applications, Université de Bordeaux, CEA, CNRS UMR 5107, Talence, 33405 France
*andre.mysyrowicz@ensta-paristech.fr

We propose a new mechanism to explain the origin of optical gain in the transitions between excited and ground state of the ionized nitrogen molecule following irradiation of neutral nitrogen molecules with an intense ultra short laser pulse. An efficient transfer of population to the excited state is achieved via field-induced multiple recollisions. We show that the proposed excitation mechanism must lead to a super-radiant emission, a feature that we confirm experimentally.

When an intense femtosecond laser pulse propagates in air, a column of weakly ionized plasma is formed. This long thin plasma column left in the wake of the pulse arises mainly from a dynamic competition between two effects: self-focusing of the laser beam and the defocusing effect that occurs when the collapsing laser pulse acquires high enough intensity to ionize air molecules [1, 2]. The plasma column emits a characteristic UV luminescence consisting of sharp lines from excited $N_2$ and $N_2^+$ molecules [3, 4]. As shown recently, these excited nitrogen molecules, both neutral and ionized, give rise to a mirror-less lasing effect [5-18]. Lasing in neutral molecules is now understood. The scheme is the same as in the traditional nitrogen laser. Population inversion is due to impact excitation of neutral nitrogen molecules by energetic free electrons produced by the intense laser pulse [9-11]. Free electrons with a sufficient energy for impact excitation are produced with circularly polarized laser pulses. On the other hand, lasing in the ionized system remains mysterious. Optical amplification occurs for transitions between the second excited state $B^2\Sigma_u^+$ and the ground state $X^2\Sigma_g^+$ of the singly ionized nitrogen molecule [12-18] (see Fig. 1(a)). Observation of optical amplification requires seeding, the injection of light at a wavelength that corresponds to a transition between vibrational sublevels of states $B^2\Sigma_u^+$ and $X^2\Sigma_g^+$. The seed can be external or self generated by the laser pulse during its propagation.

Examples of externally and self-seeded amplification are given in Fig. 1(b) for a linearly polarized pump laser pulse of 50 fs duration at 800 nm. The spectrum of the forward lasing emission is measured by a fiber spectrometer (Ocean Optics HR 4000), after removal of the pump light at 800 nm with appropriate color filters. Line at 391 nm corresponds to the transition $B^2\Sigma_u^+(v = 0) \rightarrow X^2\Sigma_g^+(v' = 0)$ where $v$ and $v'$ denotes the vibrational levels of the excited and ground ionic states. This forward lasing at 391 nm exhibits an obvious threshold with pump laser energy, as presented in Fig. 1(c). Measurement of the gain at 391 nm as a function of delay between the pump and seed pulse is shown in Fig. 2. The optical gain is achieved during the femtosecond laser pulse and lasts for several tens of picoseconds. Recently, self-seeded emission at 428 nm with several microjoules of energy per pulse was obtained in the forward direction by simply launching a multi-TW femtosecond laser pulse at 800 nm in air [16]. The seed was provided by the broadband continuum self-generated during propagation of the laser pulse. The stimulated emission corresponds to the transition $B^2\Sigma_u^+(v = 0) \rightarrow X^2\Sigma_g^+(v' = 1)$. The emission peak power was in the MW range, opening the exciting prospect to form with relative ease a powerful UV laser in the sky.

Several scenarios have been discussed to explain the origin of this optical gain, none totally convincing. One can immediately exclude a wave-mixing process involving pump and seed pulses, because a retarded seed still gives rise to the intense stimulated emission. A population inversion based on a pump induced population transfer from the neutral molecule to both ionized states followed by a collision-induced faster depletion of the ground ionic state $X^2\Sigma_g^+$ than of the excited state $B^2\Sigma_u^+$ can also be dismissed because it would require an inversion build-up time on the order of the collision time ( ~ ps) [18]. It is therefore incompatible with the observed gain that is realized within 50 fs. A recent proposal suggests that population inversion is established by depletion of state $X^2\Sigma_g^+$ through an efficient multi-photon **radiative** transfer to state $B^2\Sigma_u^+$ [18]. However, a radiative transfer is a reversible process which should not lead to population inversion. Still another scenario invokes a transient inversion between rotational packets of the ground and excited ion states [19]. Revivals of rotational wave packets of molecules are certainly pertinent in explaining the surge of gain at characteristic delays of the seed pulse [14], but do not address the gain observed between these revivals. The fundamental question remains: how is optical gain achieved in this system?

A clue to the answer of this question is given by the results of amplification as a function of the pump laser ellipticity, shown in Fig. 3(a). The amplification is strongest for a nearly linearly polarized pulse, and it is suppressed as soon as the degree of ellipticity $\varepsilon$ reaches 0.3. There is a striking analogy with the behavior of high order harmonics generation (HHG) [20], which we measured under identical experimental conditions (see Fig. 3(b)). The effect of ellipticity in high harmonic generation is well understood. A semi-classical model predicts the main features with a remarkable success [21, 22]. The high harmonic generation process is divided in three successive steps: tunnel ionization, motion of the electron wave packet in the strong laser field and recombination on the parent ion. With a circularly polarized light pulse, the returning free electron wave function never overlaps with the parent ion and therefore cannot transform its kinetic energy in the form of high harmonics. The same semi-classical model also describes non sequential double ionization (NSDI) and molecular fragmentation [23, 24]. A returning electron having acquired sufficient energy in the laser field can, upon impact, eject a second electron from the atom or molecule, leaving it in a doubly charged state or break the molecular bonding. Again, in most cases this does not happen with a circularly polarized light.

Applying the same semi-classical model to our case, we interpret the ellipticity dependence of the gain as being due to a **non-radiative** transfer of ion population from $X^2\Sigma_g^+$ to $B^2\Sigma_u^+$ state via laser field-assisted recollision. In each elementary act, an electron in the presence of the intense pump laser is removed from the outer orbital of the neutral nitrogen molecule, accelerated and then is driven back by the laser field to the parent molecular ion where it collides inelastically with an inner orbital electron. During this event, if the impacting electron has a sufficient energy, there is a probability to transfer the inner orbital electron to the outer orbital of the molecular ion. A free electron and an excited molecular ion are left after the event.

There are specific characteristics for this type of population transfer. First, in a manner similar to HHG and NSDI, this excitation process via electron recollision should exhibit a strong dependence upon pump laser polarization. It can only occur for returning free electrons with kinetic energy above 3.17 eV, the threshold energy required to bring a molecular ion in the excited state $B^2\Sigma_u^+$. This condition is easily met in our experiments. Inside filaments and/or in our low pressure experiments, the intensity reaches or even exceeds a level of $1.5 \times 10^{14}$ W/cm², which corresponds to a ponderomotive potential (average electron kinetic energy) $U_p = e^2 E_0^2 / 4 m_e \omega_0^2 = 9.1$ eV [4]. All returning electrons with kinetic energy exceeding the threshold can contribute, since the extra energy is carried by the free electron left after the process. A second characteristic of this process is its irreversibility. Ions remain stored in the excited $B^2\Sigma_u^+$ state because the reverse process (collision assisted down-transition from $B^2\Sigma_u^+$ to $X^2\Sigma_g^+$) requires the presence of an external electron, a very unlikely event during the short laser pulse duration. Furthermore, the process is parametric in the sense that the probability of transfer from $X^2\Sigma_g^+$ to $B^2\Sigma_u^+$ state can repeat at successive optical cycles of the pump pulse. The frequency of recollision in this multiple impact excitation process is crucial. To understand this point, consider the excitation of a pendulum by a single impact event. Except for a very strong impact, it will lead to small oscillations of the pendulum around its rest point. Consider now multiple impacts. If the time delay between successive percussions corresponds to a period of oscillation of the pendulum or an integer multiple of it, a large oscillation amplitude proportional to the number of impacts can be obtained because the effects of successive impacts add coherently. Translating this to our quantum case, the probability amplitude of exciting level B with a single collision can be small. However, multiple recollisions at a rate commensurable with the transition frequency can lead to a high degree of excitation, because the transition probability amplitudes add coherently. Finally, an important characteristic of this process is the special nature of the subsequent emission. Indeed, at the end of the pump pulse, the polarizations of all excited molecules are locked to a common electronic phase imposed by the pump field via the multiple attosecond recollisions. The ensuing macroscopic dipole moment leads to superradiance, the collective emission of the ensemble of excited molecular ions [25]. The emission intensity must then be proportional to the square of the number of emitters [25].

The amplitude $a_{XB}$ of the transition X → B can be evaluated according to the first order non-stationary perturbation of the Schrödinger equation:

$$i\frac{da_{XB}}{dt} = V_{XB} e^{-i\omega_{BX} t} \tag{1}$$

where $V_{XB}$ is the matrix element of the ion transition X → B in a collision with an electron having energy $\varepsilon_e > \hbar\omega_{BX}$. In our case, this excitation is produced by an electron ionized at the time $t = t_0$ (near the laser pulse maximum) and oscillating around the parent ion in the laser field. As the electron is passing around the ion many times, the matrix element can be presented as a sum of instantaneous perturbations at the moments of collisions $t_s$, that is, $V_{XB} = \sum_s V_1(\varepsilon_e)\delta(t-t_s)$, where index $s$ is an integer numbering the subsequent collisions and $\delta$ is the Dirac delta-function. The matrix element $V_1$ is, in general, a function of the electron energy. For simplicity, we consider it here as a step function $V_1 = V_0 H(\varepsilon_e - \varepsilon_{BX})$, where $H(x)$ is the Heaviside step function. Then $V_0^2$ can be considered as the probability of ion excitation in a single collision with an electron. For estimates we adopt here a value of the matrix element $V_0 = 0.1$, which corresponds to the typical ratio between the probabilities of inelastic and elastic collisions in a gas [26]. Then by integrating Eq. (1) over time one obtains the following expression for the transition amplitude

$$a_{XB} = -iV_0 \sum_s e^{-i\omega_{BX} t_s}, \qquad (2)$$

where the summation is taken over all collisions where the electron energy verifies the condition $\varepsilon_e(t_s) > \hbar\omega_{BX}$. We calculated that for conditions close to the experiment, about 22 periodic recollisions occur involving electrons of sufficient excitation energy. (The details on electron dynamics are provided in the supplementary material). As the electron motion is defined by the laser field, the subsequent recollisions with the pair and impair index $s$ are separated by the laser period for electrons born near the laser field maximum. Consequently, the expression (2) presents a sum of two geometrical series. A constructive interference of the transition amplitude between consecutive collisions takes place if the frequency detuning between the transition frequency and the second harmonic of the laser is sufficiently small, $\Delta = \omega_{BX}/2\omega_0 - 1 \ll 1$. For a sufficiently large number of recollisions and neglecting the correlation between the pair and impair events the total probability of excitation $W_{XB} = |a_{XB}|^2$ can be evaluated as

$$W_{XB} \cong \frac{V_0^2}{2(\sin 2\pi\Delta)^2}. \qquad (3)$$

The details of the derivation can be found in the supplementary material. With $\Delta \sim 0.02$, as in our experiment, the probability of finding the ionized nitrogen molecule in state $B^2\Sigma_u^+$ at the end of the pulse is essentially unity for an electron born in the appropriate phase interval of the pump pulse carrier wave.

It would be illusory to push the one-dimensional semi-classical model further and quantify the population transfer more precisely. A proper theory should include the dispersion of the free electron wave function, the Coulomb attraction that acts as a refocusing mechanism especially effective after multiple recollisions, the effect of laser intensity clamping in a filament, the effect of the Lorentz force and the competition between recollision and other processes such as high harmonic generation and double ionization.

Nevertheless, it is possible to verify experimentally a prediction of our model, namely the super-radiant nature of the transition. It is well known that the emission delay and pulse duration of a superradiance scales like $N^{-1}$, and the radiation power scales like $N^2$, where $N$ is the number of emitters, itself directly proportional to the gas pressure. We therefore measured the temporal profile of the 391 nm emission for increasing gas pressures, in the presence of a constant external seed pulse. The results are presented in Fig. 4(a). The emission delay, pulse duration, as well as the peak power of the radiation are presented in Fig. 4(b) together with best fits. The overall agreement between the experimental scaling and the theoretical expectation confirms the superradiance nature of the 391 nm lasing. We attribute the slight deviation from theoretical expectations to the presence of the self-generated seed pulse that increases with gas pressure, and therefore accelerates the super-radiation process even further. We also note that a signature of superradiance has been reported recently by G. Li and co-workers [13].

In conclusion, we have proposed a scheme explaining the optical amplification of ionic origin occurring in air or in pure nitrogen molecular gas irradiated with intense femtosecond laser pulses. At sufficiently high laser intensities, a sequence of field-assisted inelastic recollisions takes place, promoting nitrogen molecules to the excited ionic state and possibly preparing an inverted population with respect to the ground ionic state. This scenario explains naturally the ultrafast gain built-up (since the inelastic recollisions occur within the pump pulse), the sharp dependence of gain upon pump polarization, as well as the superradiance nature of the emission.

**Acknowledgement**

We acknowledge fruitful discussions with P. Mora of Ecole Polytechnique, and P. Corkum and M. Spanner from NRC in Ottawa. Yi Liu is grateful for Prof. H. B. Jiang and C. Wu of Peking University for stimulating discussion and acknowledges the supports of the open research funds of State Key Laboratory for Artificial Microstrucutre and Mesoscopic Physics of Peking University.

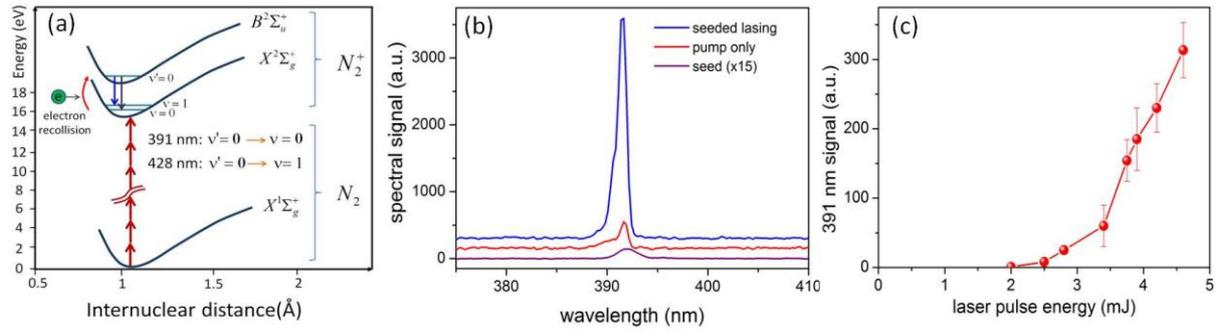

Figure 1. (a) Schematic diagram of relevant energy levels of nitrogen molecules and its ions. (b) Amplification of the seed pulse inside the plasma filament. The femtosecond pump laser pulse is focused by a convex lens of *f* = 400 mm in pure nitrogen at a pressure of 30 mbar. The seed pulse around 391 nm is generated by second harmonic generation in a thin BBO crystal on the second arm of a Mach-Zehnder interferometer. The weak seed pulse is combined with the 800 nm pump pulse with a dichromatic mirror. The seed pulse is also focused by an *f* = 400 mm lens installed before the dichromatic mirror. The spatial overlap and temporal delay of the pump and seed pulses is carefully controlled. The pump pulse energy is 3 mJ. Self-seeded stimulated emission is responsible for the weaker signal observed with pump only. (c) The self-seeded 391 nm signal as a function of pump laser energy in 150 mbar nitrogen.

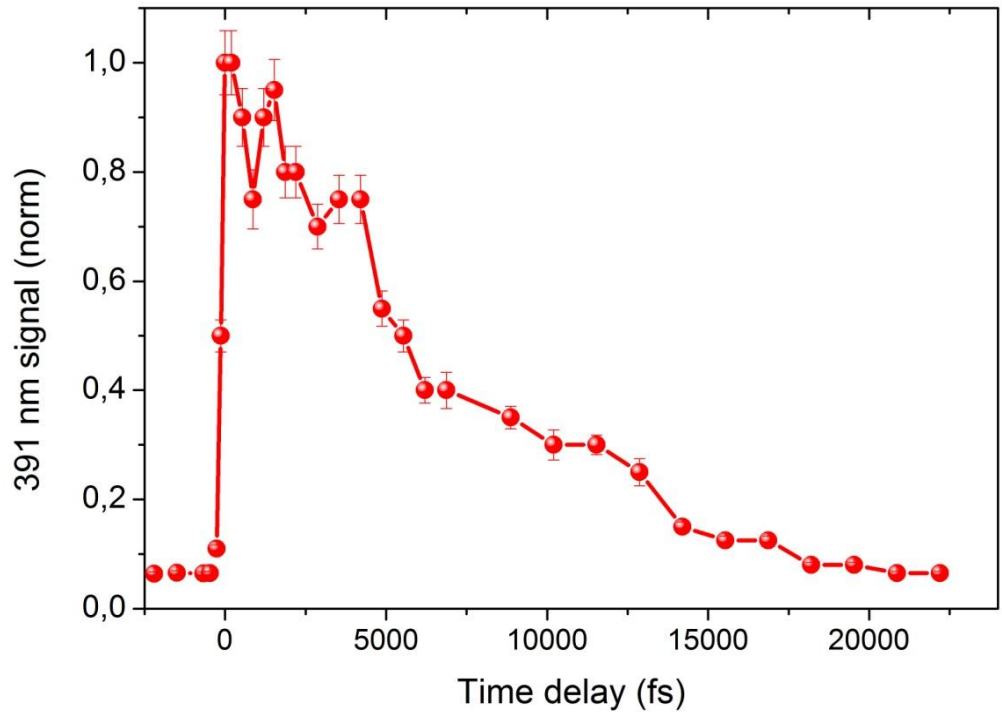

Figure 2. Temporal evolution of the optical gain at 391 nm measured in 10 mbar nitrogen gas as a function of delay between pump and seed pulse. Pump pulse energy is 3 mJ. Seed pulse characteristics are as in Figure 1.

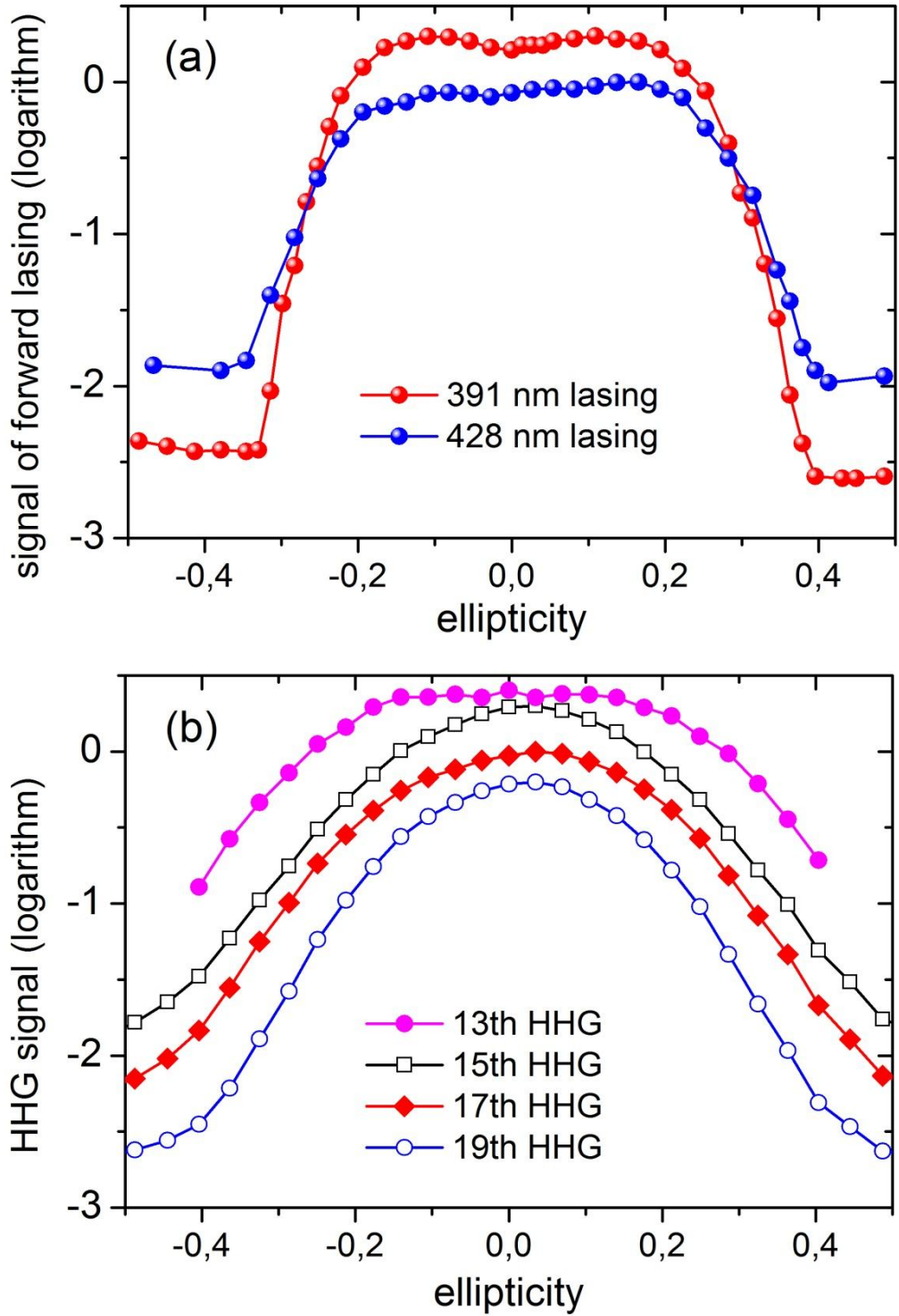

Figure 3. (a) 391 nm and 428 nm lasing emission as a function of the ellipticity of the pump pulses at 800 nm. The gas pressures were 45 mbar and 300 mbar respectively. (b) Dependence of the high-order harmonic yield in nitrogen gas as a function of the laser ellipticity. The experiments were performed in a 15 mm long gas cell with two 150 μm holes on the entrance and exit surfaces of the gas cell, which were drilled by the laser pulse itself on the 100 μm thick aluminum foils windows. Each data point was an average over 10 thousand laser shots. The nitrogen gas pressure was about 65 mbar and the pump pulse energy was 3.8 mJ.

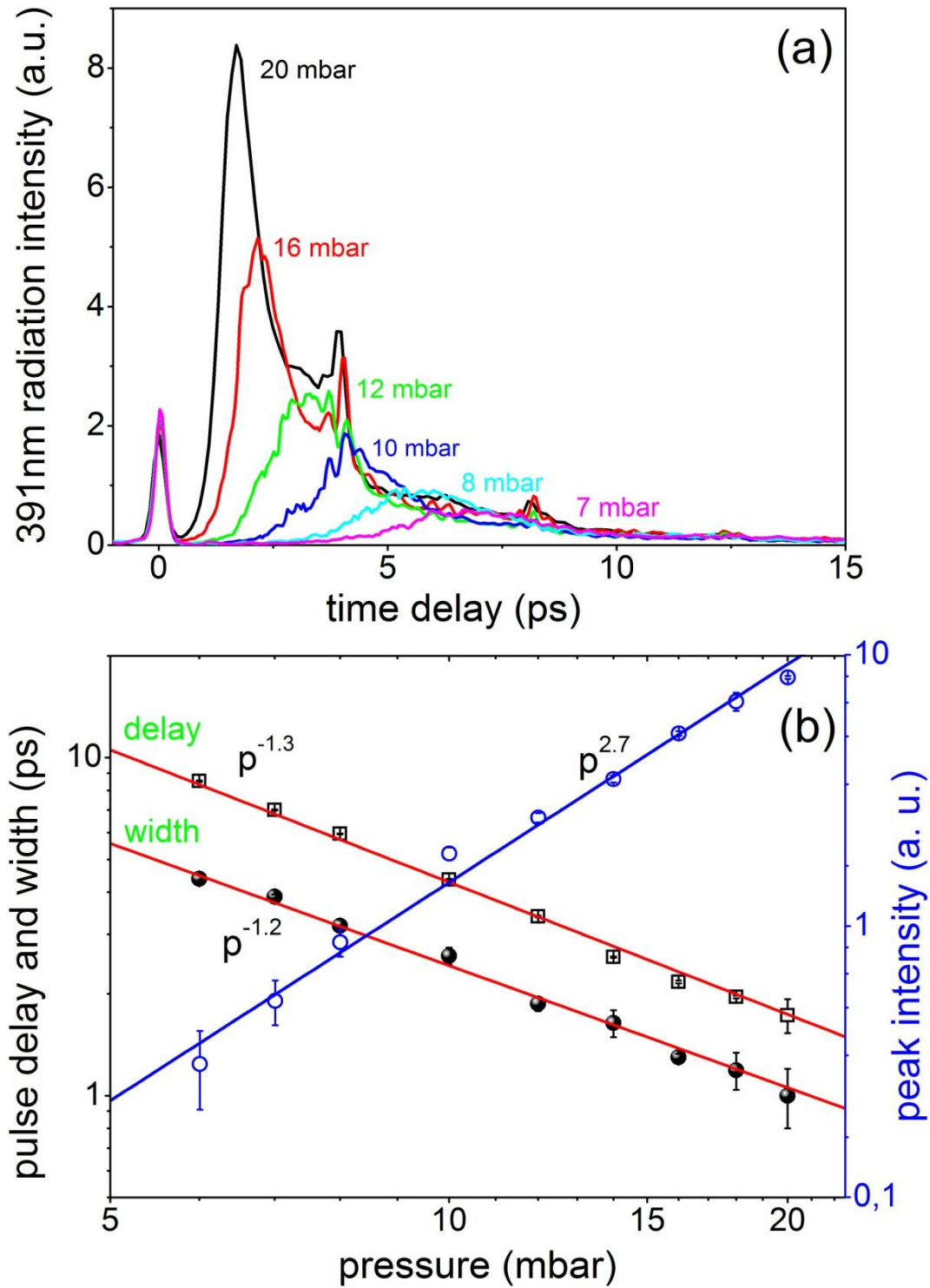

Fig. 4. (a) Temporal profile of the 391 nm forward radiation measured for different nitrogen pressure in the presence of a constant seed pulse. The narrow peaks at zero delay are the measured seed pulse. These temporal profiles are measured by frequency mixing the 391 nm radiation and a constant 800 nm pulse inside a sum frequency generation (SFG) BBO crystal. The SFG signal at 263 nm is recorded as a function of relative delay between the two pulses. (b) The pulse delay, pulse width, and the 391 nm peak intensity are presented for increasing nitrogen pressure.

# Supplementary material

## 1. Modeling and calculation of the multiple electron recollision

The electron recollision dynamics was calculated from the solution of the Newton equations for the ionized electron moving in the laser electric field:

$$\frac{dx_e}{dt} = v_e, \quad \frac{dv_e}{dt} = -\frac{e}{m_e} E_x(t) \cos \varphi(t)$$

where $x_e$ and $v_e$ are the coordinate and the velocity of electron, $e$ and $m_e$ are the elementary charge and electron mass respectively, $E_x(t) = E_0 \sin(\pi t / 2t_{las})$ is the laser amplitude envelope with the maximum amplitude $E_0$ and the pulse duration $t_{las}$. The phase

$$\varphi(t) = \omega_0 t + \frac{1}{4} C (t/t_{las} - 1)^2$$

depends on the laser frequency $\omega_0$ and the chirp $C = \sqrt{(t_{las}/t_{min})^2 - 1}$ with $t_{min}$ being the shortest laser pulse duration ~50 fs in our case. These equations were solved numerically starting from the electron birth time $t_0$ and up to the end of the laser pulse. The electron is born at the parent ion position $x_e(t_0) = 0$ with zero velocity $v_e(t_0) = 0$ and with a probability following from the theory of tunnel ionization. The laser peak intensity is $1.5 \times 10^{14}$ W/cm² at the wavelength of 800 nm. We calculated the time moments $t_s$ when the electron returns to the origin and the corresponding values of its kinetic energy $\varepsilon_e(t_s)$.

Figure S1 shows an example of the electron trajectory and the values of the electron energy at the recollision moments. Here we chose an appropriate birth phase $t_0 = 20.979 T_{las}$, where $T_{las} = 2\pi/\omega_0$ is the laser period, a chirp factor $C = 2$ and a laser pulse of 43 optical cycles. This chirp was introduced because we noticed in the experiments that the maximum 391 nm stimulated emission occurs when the laser pulse has a positive chirp $C = 2$. The electron stays near the parent ion during about 22 laser periods (see panel a) and then gradually drifts away. The electron collides with the ion two times per laser period. It has an energy sufficient for ion excitation ($0.3 U_p = 3.17$ eV in our case) from the 23[rd] to the 55[th] laser period.

Using the semi-classical model of HHG [21, 22], we have calculated the probability of ion excitation by an electron born at different instants near the maximum of the pump laser field according to Eq. (1). The total amplitude of the transition in the state B is given by Eq. (2) assuming that the collision time is much shorter than the laser period. The excitation probability has been calculated from this expression numerically. For estimates, it can be further simplified by taking into account the fact that the time interval between the subsequent pair and impair collisions is approximately equal to the laser period (see Fig. S1(b)). Consequently, the pair and impair terms in Eq. (2) present a geometrical series that can be evaluated in the limit $s \gg 1$:

$$a_{XB} = -iV_0 \frac{e^{-i\omega_{BX}t_1}}{1-e^{-i\omega_{BX}T_{las}}} - iV_0 \frac{e^{-i\omega_{BX}t_2}}{1-e^{-i\omega_{BX}T_{las}}}.$$

Assuming that two terms in the right hand side are incoherent, one readily obtains expression (3). The analytical estimate is in good agreement with the numerical evaluation. The total probability of the ion excitation was calculated by integrating the excitation probability over the ionization time $t_0$ taking into account the ionization probability. Typically the ionization takes place twice the laser period in the time interval $\Delta t_0 \sim 0.05 T_{las}$ during 5-10 laser periods near the pulse maximum. It is important to mention that all ions are excited in approximately the same phase, which creates the favorable condition for the stimulated emission after the end of the laser pulse.

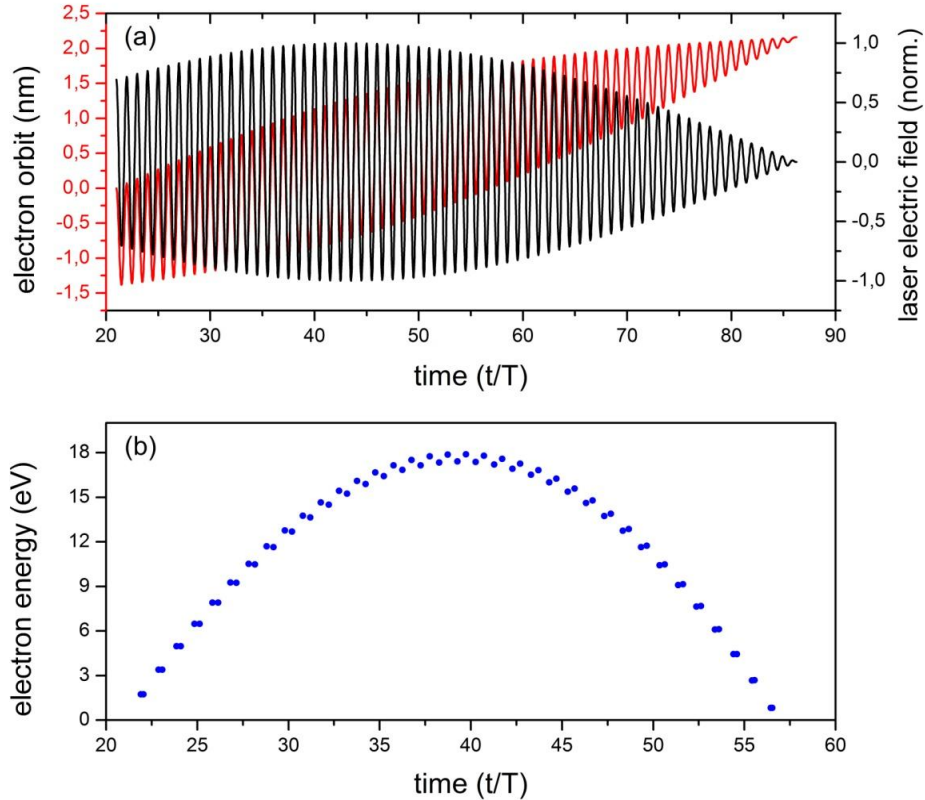

Figure S1. (a) Laser pulse electric field (black) and the electron position (red) as a function of time. The laser pulse has a linear chirp $C = 2$, and a pulse duration $t_{las} = 115$ fs. (b) Electron energy at the moment of recollisions as a function of time. The time is measured in laser periods. Laser parameters are given in the text. The electron is born at the time $t_0 = 20.979$ laser periods, that is 0.021 period from the laser field maximum.

## 2. Experimental method for the measurement of superradiance

To measure the temporal evolution of the forward 391 nm emission, we employed the cross-correlation method based on the sum-frequency-generation of the forward 391 nm emission and a weak 800 nm probe pulse in a BBO crystal. One typical spectra of the sum frequency generation process is presented in Fig. S2, where the probe pulse at 800 nm, the lasing emission at 391 nm, and the generated 263 nm signal are all visible. The SFG signal at 263 nm is recorded as a function of relative delay between the 391 nm radiation and the 800nm probe pulse.

We have performed measurements concerning the forward 391 nm in both the self-seeded regime and the externally seeded regime. As the externally seeded 391 nm signal is much intense compared to that of self-seeded one (Fig. 1 (b)), the signal-to-noise ratio of the cross-correlation measurement is better in the presence of the external seeding pulse. The results are presented in Fig. 4(a). In these experiments, the external seeding pulse at 391 nm is nearly synchronized with the pump 800 nm laser, because the maximum optical gain is achieved when the pump and seeding pulse overlap with each other (Fig. 2).

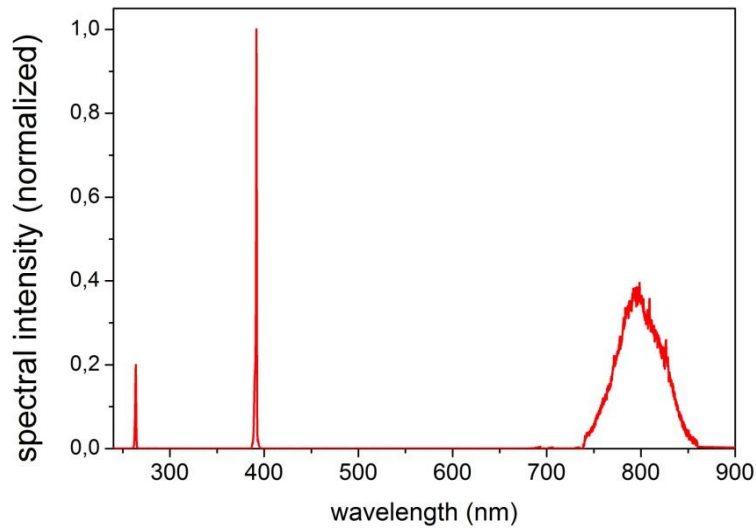

Fig. S2. Spectrum measurement for the sum frequency generation process. The optical signal at 263 nm is generated by the mixing of the forward 391 nm radiation and the 800 nm probe pulse.